\documentstyle[11pt,paspconf,epsf]{article}

\begin{document}

\title{A 2.14 ms Candidate Optical Pulsar in SN1987A}

\author{J. Middleditch\altaffilmark{1,2}}
\affil{Modeling, Algorithms, and Informatics, CCS-3, MS B256, 
Los Alamos National Laboratory, Los Alamos, NM 87545, jon@lanl.gov}

\author{J. A. Kristian\altaffilmark{1,3} and W. E. Kunkel\altaffilmark{1}}
\affil{Observatories of the Carnegie Institution of Washington, 813
Santa Barbara St., Pasadena, CA 91101}

\author{K. M. Hill \& R. D. Watson}
\affil{School of Mathematics and Physics, 
University of Tasmania, Hobart, Tasmania, Australia}

\altaffiltext{1}{Visiting Astronomer, Cerro Tololo Inter-American Observatory. 
CTIO is operated by AURA, Inc.\ under cooperative agreement with the National
Science Foundation} 
\altaffiltext{2}{Other co-authors include:~ R.~Lucinio, 
Independent Consultant and Division of 
Geological \& Planetary Sciences, 170-25,
Caltech, Pasadena, CA 91125; J.N.~Imamura, Institute for 
Theoretical Science, University of Oregon, Eugene, OR
97403; T.Y.~Steiman-Cameron, MS245-3,
S.M.~Ransom, Harvard-Smithsonian
Center for Astrophysics, 60 Garden St., Cambridge, MA 02138;
A.~Shearer \& R.~Butler, Institute for Theoretical Computing, 
\& M.~Redfern, Physics Department, University 
College, Galway, Ireland; \& A.C.~Danks, NASA Goddard SFC,
Code 683.0, Bldg.~21, Greenbelt, MD 20771.}
\altaffiltext{3}{We lost our esteemed colleague, Dr. Kristian, in June 1996.} 

\begin{abstract}

We have monitored Supernova 1987A in optical/near-infrared bands
from a few weeks following its birth until the present time 
in order to search for a pulsar remnant.
We have found an apparent pattern of emission near the
frequency of 467.5 Hz -- a 2.14 ms pulsar candidate, first
detected in data taken on the remnant at the Las Campanas
Observatory (LCO) 2.5-m Dupont telescope during 14-16 Feb.~1992 UT.
We detected further signals near the 2.14 ms period on numerous
occasions over the next four years in data taken with a variety of
telescopes, data systems and detectors, at a number of ground- and
space-based observatories.  The sequence of detections of
this signal from Feb.~`92 through August `93, prior to its 
apparent subsequent fading, is highly improbable (${<10^{-10}}$
for any noise source).  We also find
evidence for modulation of the 2.14 ms period with a
$\sim$1,000 s period which, when taken with the high spindown
of the source (2-3$\times$10$^{-10}$ Hz/s),
is consistent with precession and spindown via
gravitational radiation of a neutron star with a non-axisymmetric
oblateness of $\sim$10$^{-6}$,  and an implied gravitational luminosity
exceeding that of the Crab Nebula pulsar by an order of magnitude.  
\end{abstract}


\section{Introduction}

We present evidence herein for the presence of an unique, 2.14-ms pulsar,
which appears to precess and whose spindown is apparently dominated by 
gravitational radiation, but one which has also been very difficult to
detect over the past three years due to its faintness and 
the complicated nature of its signal.  
\begin{figure}
\plotfiddle{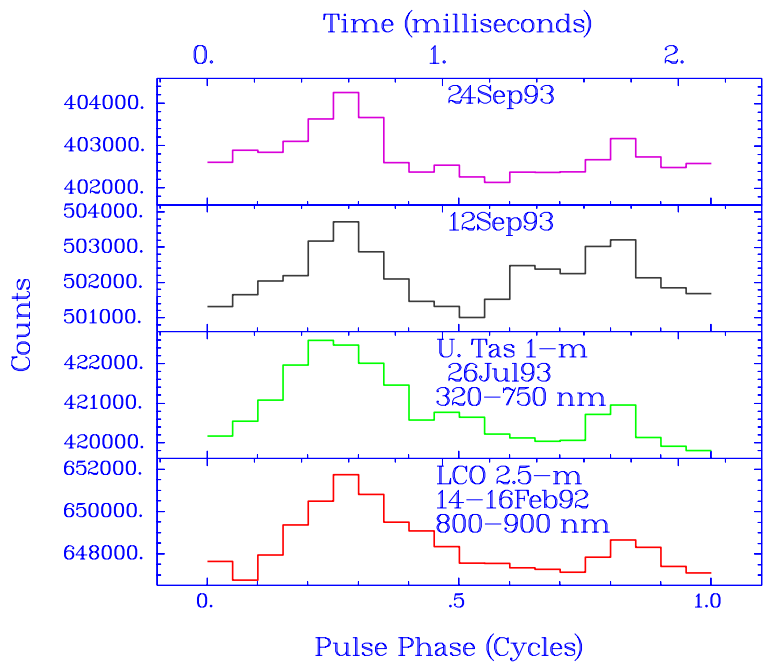}{3.0in}{-90}{50}{50}{-162}{277}
\caption{The pulse profiles for the 2.14 ms periodicity detected 
on UT 26 July, 12 Sep.~and 24 Sep.~`93 with the 1-m telescope are 
plotted against the UT 14-16 Feb.~`92 detection with the 
LCO 2.5-m telescope.} \label{fig-1}
\end{figure}
A more detailed and exhaustive treatment of this candidate 
and upper limits for any other signal are given elsewhere
(Middleditch et al. 2000).

\section{Observations}

The observations discussed in this work were made 
between Feb.~`92 and Feb.~`96, using the LCO 2.5-m, CTIO 4-m,
U.~Tasmania 1-m, and the ESO NTT (3.5-m) telescopes. 
In addition, archival data from the HST on SN1987A was
also used.  A GaAs tube with an 800 nm longpass
filter were used for the first year (ibid.).

\section{Discovery of the 2.14 ms Signal}

\subsection{LCO Feb.~14-16, 1992}

We first detected the 2.14 ms periodicity in data taken at the LCO
2.5-m Dupont telescope during 1992 Feb.~14-16 UT.
The 2nd harmonic phase was twice that of the
fundamental, a sure sign of a ``peaked'' pulse profile,
which turned out to have a noise probability of
$\sim e^{-27.45}$ when actually computed. This profile 
is shown in Fig.~1, folded into 20 phase bins per cycle, 
together with results from the U.~Tas 1-m
which will be discussed in part below.
The statistics of the pulse profiles are derived from
folding into 11 phase bins.

\subsection{LCO February 1993 -- The Rosetta Stone}

We continued to observe SN1987A with the 800 nm longpass filter
in the year following the initial appearance of the 2.14-ms signal.  
A targeted search was made using the sum of the complex 
Fourier amplitudes, a(f), near the $\sim$467.5 Hz fundamental frequency,
f, with those near the $\sim$935 Hz second harmonic, a(2f),
computed in the following way:~
$a_{sum}=(~\Vert a(f)\Vert+e^{-i({\phi}(2f)-2{\phi}(f))}\Vert a(2f)\Vert)/\sqrt{2}$
where ${\phi}(f)$ is the phase of $a(f)$, and ${\phi}(2f)$ is the
phase of $a(2f)$.  
\begin{figure}
\plotfiddle{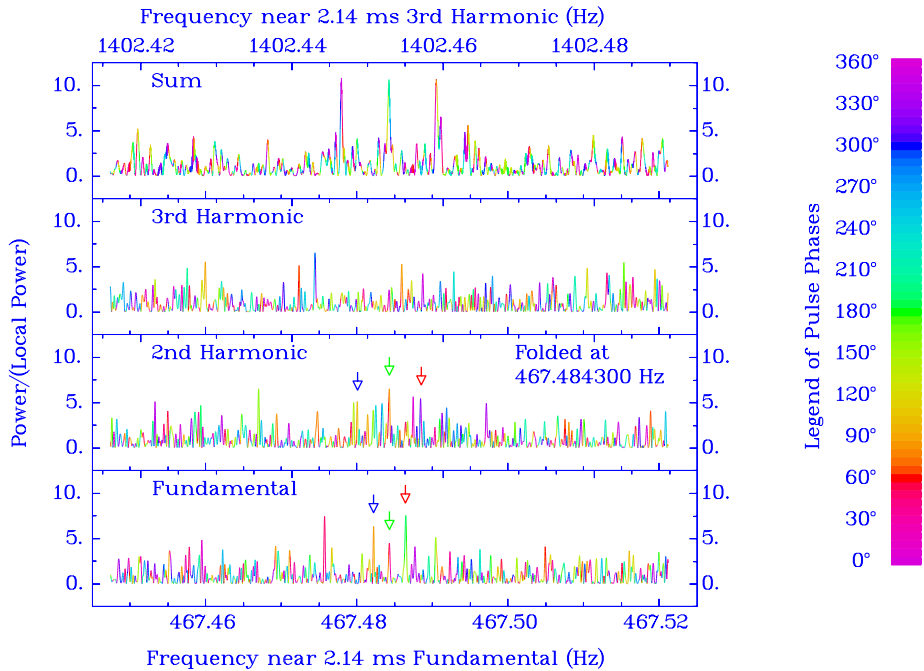}{3.0in}{-90}{50}{50}{-202}{277}
\caption{(Lower three frames) The Fourier power spectra
for frequency regions near 467.4843 Hz and its first
two higher harmonics from data taken at the LCO 2.5-m
on Feb.~6, 1993.  (Top) The sum spectrum of the
fundamental and 2nd harmonic (see $\S$3.2).}\label{fig-2}
\end{figure}

On the night of 6 Feb.~UT 1993, in a run on the LCO 2.5-m which lasted
$\sim$80 minutes before high humidity forced an early termination
(conditions were otherwise perfect), an unusual pattern of power
appeared in the sum of the Fourier spectra from frequency regions
encompassing 467.4843 Hz (close to the extrapolation of the Feb.~`92 frequency)
and twice this value (Fig.~2). 

The three high peaks in the sum power spectrum,
are, to within errors, evenly spaced by 0.00214 Hz,
modulo the 467.5 Hz fundamental frequency,
and immediately suggest a periodic modulation in
the phase/frequency and/or amplitude of the 2.14 ms
signal with a period of $\sim$467 s.  
The 467.48429 Hz frequency of the central peak
(1/3 of the frequency of the top scale)
also indicates a {\it mean} spindown for the pulsar
implied by the 2.14 ms signal
for the $\sim$1 year interval between Feb.~`92 and Feb.~`93
of about -3$\times$10$^{-10}$ Hz/s.

The probability of noise producing three such peaks 
in the sum spectrum,
each exceeding 10 times mean power, is less than $10^{-5}$,
even considering the rest of the data from Feb.~`93 and
Nov.~`92 (Middleditch et al. 2000).  
Further modulation structure of the individual Fourier
power peaks of the fundamental and 2nd harmonic lead
to a further, very conservative, reduction in this
probability by a factor of 100, in addition to the 
realization that the period of modulation was actually
twice 467.5 s, or 935 s.  Thus the probability fell
to 10$^{-7}$.
\begin{figure}
\plotfiddle{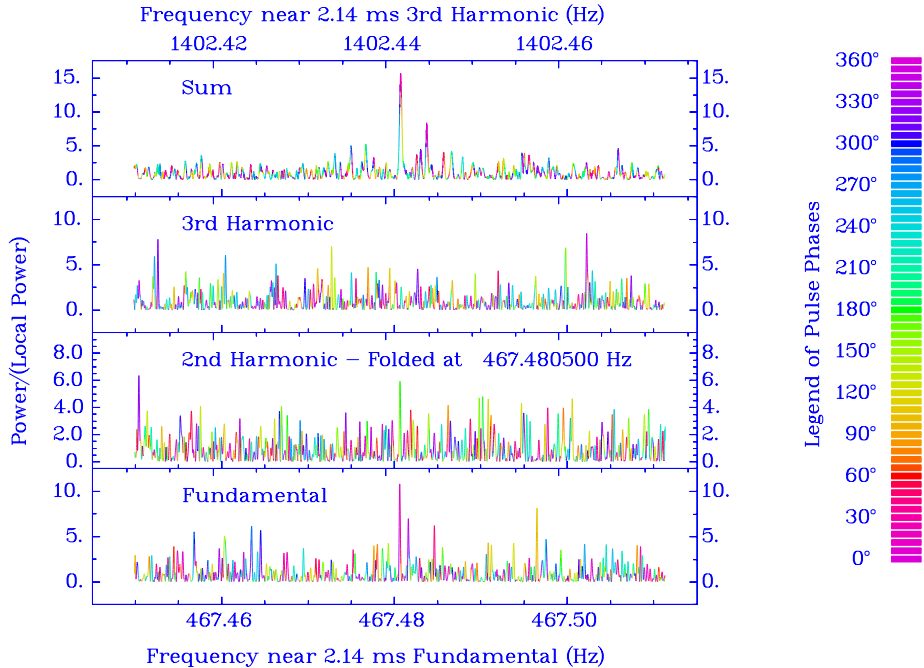}{3.0in}{-90}{50}{50}{-202}{277}
\caption{(Lower three frames) The Fourier power spectra for 
frequency regions near 467.48056 Hz and its first two higher
harmonics (the 2nd near 935 Hz and the third near 1402.5 Hz) from data
taken at the U.~of Tasmania 1-m during July 26 UT, 1993. (Top frame) 
The sum spectrum of frequencies near the fundamental and 2nd harmonic.
The peak in the sum spectrum near
1402.4417/3 = 467.48056 Hz is significant above the five sigma level
($\sim$1:6,500,000).}\label{fig-3}
\end{figure}

The remainder of the data from LCO during 1992 and Feb.~1993 was
searched for the 2.14 ms signal and $\sim$1,000 s modulation,
and evidence was found for this signal and modulation in data taken on
the other nights at LCO and CTIO (the 3rd, 5th, 7th, 
11th, and 12th of Feb.~`93)
and in two of three nights during Nov.~`92 (the 6th and 8th).  In
addition, a moderately strong and consistent 1430 s modulation
was found in the data from Feb.~`92.

\section{Observations from Tasmania, 1993}

By the time the pattern of the 2.14 ms signal was
beginning to be discerned, the regular `92/`93 observing 
season was over.  In order to continue observations during 
an interval in which the signal seemed reasonably consistent, 
observations had to be made from farther south.
Thus the search continued (in the broad S20 band) 
with the 1-m telescope of the
University of Tasmania, which was sufficiently far
south to allow observations of SN1987A while under
the pole (and through at most 2.7 airmasses).

\begin{figure}
\plotfiddle{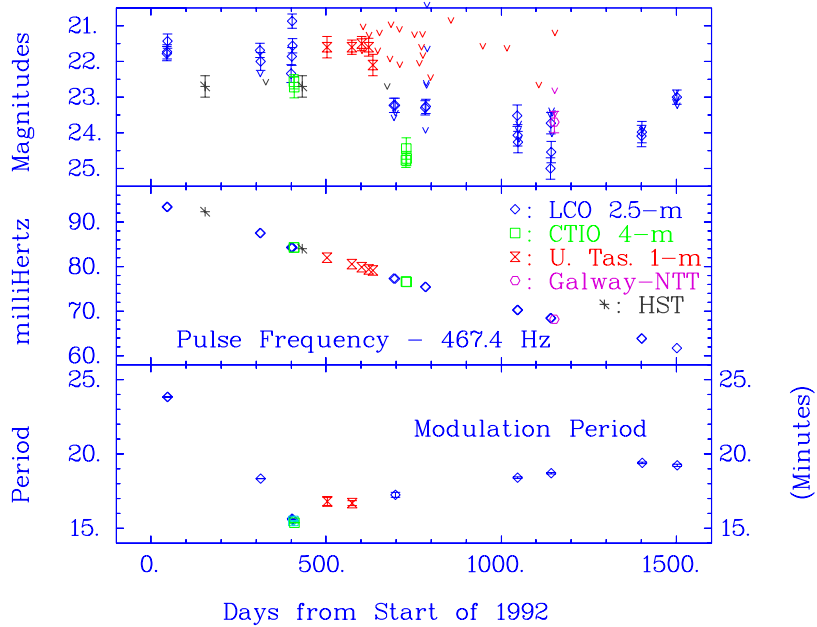}{3.0in}{-90}{50}{50}{-182}{277}
\caption{The time histories of the $\sim$1,000 s modulation period
(lower), the $\sim$ 467.5 Hz pulse frequency (middle),
and the inferred I flux (upper) for points earlier than
day 500, V + R + I composite for LCO and CTIO points
afterward, and S20 band magnitude for HST, Galway/NTT,
and U.~Tas.~points.  The `v' symbols are upper limits.}\label{fig-4}
\end{figure}

A peak in the Fourier power spectrum at 467.481992 (5) Hz was found in
the initial observation made from Tasmania on May 16, 1993 UT. 
The pulse profile had ``flat'' main and interpulses,
both of which sharpened far more than would be expected 
from noise (which would have added only relatively ``smooth'' structure)
when the obvious $\sim$1,000 s phase modulation
of the 2nd harmonic was incorporated into the pulse folding
(see Middleditch et al.~2000).

The next observation, made on July 26, 1993, has
over ten times mean power for a fundamental
frequency near 467.48056(1) Hz, and a second harmonic
with about half as much power at exactly twice this
frequency, and stands out very clearly in the sum spectrum
(Fig.~3).  The pulse profile of this signal
confirmed the Fourier spectral analysis, indicates a mean magnitude of
21.6 for the 320-750 nm S20 band, and is similar to
that of the Feb.~`92 observations (Fig.~1).

SN1987A was observed again four weeks later on 23 August
`93.  Analysis of this observation revealed
a peak in the sum of the 2nd and third harmonics,
and in addition, power in the fourth and fifth harmonics,
when folded at 467.479874(4) Hz.
Unusually high numbers (0.67$\pm$0.073 and $\pm$0.095, 
where 0.5 is expected -- Middleditch et al.~1993) for the centroids of
harmonics 2 and 5 (3 and 4 were still $>$ 0.5) led to the discovery
that most of the contribution to the
main pulse occurred during the last hour of the observation.
The pulse profile for this hour, a single sharp pulse, 
is less probable still than that of the July 26 data 
(2.44$\times$10$^{-7}$), and the peak in the 
sum spectrum produced by the 2nd and 
third harmonics for this run is even
higher than that shown in Fig.~3.
\begin{figure}
\plotfiddle{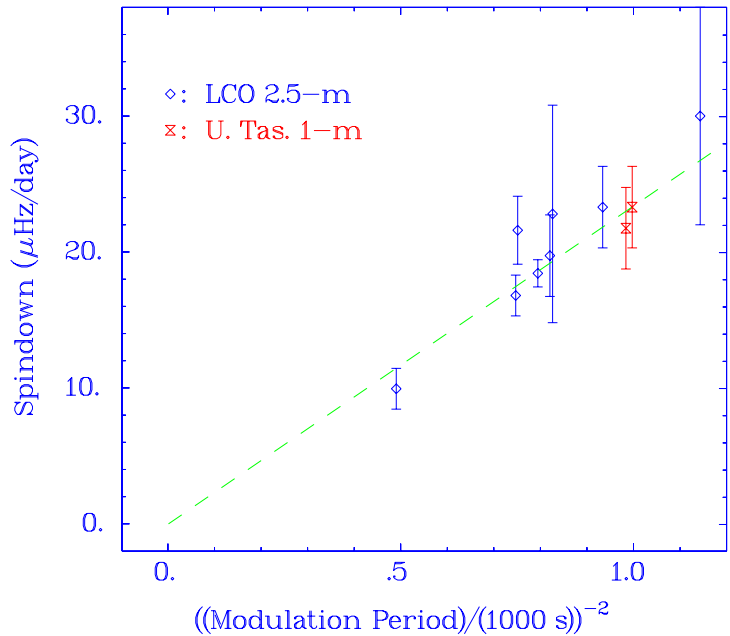}{3.0in}{-90}{50}{50}{-162}{277}
\caption{The spindowns of the 2.14 ms pulsations are
scattered against the reciprocal of the squares of their
$\sim$1,000 s periods.}\label{fig-5}
\end{figure}

Observations of SN1987A and detections of the 
2.14 ms signal continued on the big telescopes, 
in spite of the fading of the signal after Sep.~`93
(Fig.~4).  The data from all telescopes, large and
small, remain consistent.

If the apparent precession is a result of the {\it same}
non-axisymmetric oblateness (relative to the axis of rotation)  
which drives the apparent spindown, then
$\Omega_{prec} \propto {\delta I \over I} \omega_{spin}$ and 
${\partial\omega_{spin} \over \partial t} \propto ({\delta I})^2$,
where $\omega_{spin}$ is the rotational frequency of the pulsar,
$\Omega_{prec}$ is the precession frequency, I is
the moment of inertia of the (precessing part of) the
neutron star, $\delta I$ is the non-axisymmetric
contribution to I, and ${\partial\omega_{spin} \over \partial t}$ 
is the spindown.  Combining these gives,
${\partial\omega_{spin} \over \partial t} \propto \Omega_{prec}^2$
(Fig.~5).

\acknowledgements

We would like to acknowledge the Australian Academic
Research Network for network access.
JNI, TYS-C \& SMR were also visiting astronomers at
CTIO.  This work was also based in part on observations
made at the European Southern Observatory, use of the
resources of the Advanced Computing Laboratory of
Los Alamos National Laboratory, and performed 
under the auspices of the Department of Energy.

\end{document}